\documentclass[aps,pre,twocolumn,superscriptaddress,showpacs]{revtex4-1}

\usepackage{graphicx}

\usepackage{color}
\usepackage{amsmath}
\usepackage{amsfonts}
\usepackage{amssymb}
\usepackage{dcolumn}
\usepackage{float}

\begin{document}
  \title {Effect of particle polydispersity on the irreversible adsorption of fine particles on\\ patterned substrates}

  \author{J.~F.~Marques}
    \affiliation{GCEP-Centro de F{\'i}sica da Universidade do Minho, 4710-057 Braga, Portugal}

  \author{A.~B.~Lima}
    \affiliation{Universidade Federal do Tri{\^a}ngulo Mineiro, Av. Frei Paulino, 30, CEP 38025-180, Uberaba, MG, Brazil}

  \author{N.~A.~M.~Ara{\'u}jo}
    \email{nuno@ethz.ch}
    \affiliation{Computational Physics for Engineering Materials, IfB, ETH Z{\"u}rich, Schafmattstr.~6, 8093 Z{\"u}rich, Switzerland}

  \author{A.~Cadilhe}
    \email{cadilhe@lanl.gov}
    \affiliation{Theoretical Division, MSK 717, Los Alamos National Laboratory, Los Alamos, NM 87545, USA}

  \date{\today}

  \begin{abstract}
    We performed extensive Monte Carlo simulations of the irreversible adsorption of polydispersed disks inside the cells of a patterned substrate.
    The model captures relevant features of the irreversible adsorption of spherical colloidal particles on patterned substrates.
    The pattern consists of (equal) square cells, where adsorption can take place, centered at the vertices of a square lattice.
    Two independent, dimensionless parameters are required to control the geometry of the pattern, namely, the cell size and cell-cell distance, measured in terms of the average particle diameter.
    However, to describe the {\it phase diagram}, two additional dimensionless parameters, the minimum and maximum particle radii are also required.
    We find that the transition between any two adjacent regions of the {\it phase diagram} solely depends on the largest and smallest particle sizes, but not on the shape of the distribution function of the radii.
    We consider size dispersions up-to $20\%$ of the average radius using a physically motivated truncated Gaussian-size distribution, and focus on the regime where adsorbing particles do not interact with those previously adsorbed on neighboring cells to characterize the jammed state structure.
    The study generalizes previous exact relations on monodisperse particles to account for size dispersion.
    Due to the presence of the pattern, the coverage shows a non-monotonic dependence on the cell size.
    The pattern also affects the radius of adsorbed particles, where one observes preferential adsorption of smaller radii particularly at high polydispersity.
  \end{abstract}
  \maketitle

  \section{Introduction}
    Monolayer colloidal films grown by particle adsorption on substrates have a long standing interest from both scientific and technological perspectives.
    Recently, the interest has shifted towards structured films at the submicron scale with the intention to achieve the nanoscale.
    From the experimental perspective, the emphasis has been on photonic crystals, quantum dots, heterogeneous catalysts, and microarrays \cite{Kumacheva2002, Fustin2003, Zhu2004, Burda2005, Lewis2005, Oconnor2005, Yang2005}.
    In particular, there has been progress on achieving highly reproducible and regular submicron patterns \cite{Jeon1997, Chen02_2002, Kumacheva2002, Elimelech2003, Lewis2005, Joo2006, Xia2006}, and
    control of the final structure of such films is of paramount importance \cite{Chen01_2000, Zheng2002}.
    In this context, new probing techniques to experimentally follow the kinetics of the films along with novel theoretical methodologies are required.
    Specifically, theoretical studies on the influence of a pattern on the irreversible adsorption of particles have so far remained restricted to the monodisperse case \cite{Cadilhe2007, Adamczyk2008b, Araujo2008}.
    Studies on polydispersed particles have been performed on regular, i.e., non-patterned substrates \cite{Meakin1992a, Meakin1992b, Adamczyk1997, Gromenko2009a, Gromenko2009b}.
    Nonetheless, experimentally produced colloidal particles are inherently polydisperse, so there is a need to understand the effect of size dispersion on the morphology of the film.

    We partially fill the gap by presenting a study concerned with the geometrical parameters and properties of the deposition process, and specifically on the interplay of the adsorption of size dispersed particles on patterned substrates.
    The pattern consists of an array of cells centered on the vertices of a square lattice.
    Particles are allowed to deposit inside the cells, which we consider in the present study to be squares of equal size.
    To this end, particles are considered to attempt adsorption on the substrate with the same probability regardless of their position, so that their motion is not, for example, affected by the hydrodynamics of the solution \cite{Bafaluy1993, Senger1991, Senger1992, Senger1993, Kumacheva2003, Duffadar2009, Zhdanov2010}.
    Particles, which we model as disks, successfully attach to the substrate when their geometrical centers fall within a cell and also do not overlap previously adsorbed ones.
    If a particle overlaps a previously adsorbed particle or if its geometrical center does not fall within the cell region, it is assumed that the deposition attempt fails and the particle moves away from the substrate.
    Moreover, once adsorbed they cannot either diffuse on or detach from the substrate \footnote{This might not be necessarily true for certain systems involving proteins as discussed, e.g., \cite{Privman2000b, Privman1991b}.}, so we consider the ideal case of irreversible adsorption.
    In this context of accounting for the basic (geometrical) parameters, the adsorption process is well described by the random sequential adsorption (RSA) model \cite{Evans1993, Privman2000a, Privman2000b, Privman2000ed, Talbot2000, Schaaf2000, Cadilhe2007, Araujo2008}.
    Landing particles are considered to have a truncated Gaussian-size distribution with dispersions up to $20\%$ of the average radius as this represents a closer description of experimental systems.
    Finally, we focus our study on the regime where the kinetics within a cell is decoupled from the neighboring ones, i.e., the cell-cell distance is equal or larger than the diameter of the largest particle.

    The paper is organized as follows: in the next section we present the motivation for  the present study along with a description of the model.
    In Sec.~\ref{sec:results} we generalize concepts previously defined for monodisperse to polydisperse particles, while in Sec.~\ref{sec:mc.sim} we present Monte Carlo simulations characterizing the film morphology.
    Finally, in Sec.~\ref{sec:conclusion} we present our concluding remarks.

  \section{Motivation}\label{sec:model}
    We address the study of the irreversible adsorption of polydisperse colloidal particles on patterned surfaces.
    The pattern takes a particular shape repeated over the surface and in the present study we use as a template the vertices of a square lattice.
    We further assume that adsorption of particles takes place solely within these regions we term as cells.
    We adopt the case of square cells for simplicity, but depending on the experimental technique used to generate these, they can assume different shapes with stripes and circles being experimentally produced~\cite{Odom2002, Paul2003, Xia2006}.
    Though we define the pattern as regular and deterministic, one might consider relaxing such conditions and consider random patterns.
    For example, an experimental realization of such a pattern can consist of three-dimensional hexagonal inverted pyramids on In-containing and III-nitride substrates, where colloidal nanoparticles of linear dimensions in the range $5-30\mbox{nm}$ pack inside them \cite{Pereira2008}.
    Another example is the independent adsorption of colloidal particles of different sizes~\cite{Adamczyk1998, Cadilhe2004}.

    The classical linear dimensions of colloidal particles are few microns down to about a micron, but recently the trend has been to use particles of sizes well into the submicron length scale.
    Similarly, the typical geometrical cell length has fallen below the submicron, and again, with a goal to approach the nanoscale.
    Even taking into account such experimental advances, the size of the colloidal particles remains much larger than the length scale between binding sites, which are typically of the order of the linear dimensions of the substrate molecules.
    In this context, adsorption at the length scales of the cell linear dimensions can be regarded as an intrinsically off-lattice process.
    We assume that diffusion and detachment do not occur on a time scale commensurate with experimental observations, so it seems reasonable to adopt the ideal case of irreversible adsorption \cite{Voegel1985, Pefferkorn1985, Nygren1994}.
    We are primarily concerned with the effects provided by basic geometrical features of the pattern, and consequently discard other (possibly relevant) interactions beyond the excluded volume interaction.
    Since particles only adsorb inside cells of the substrate, and do not interact between themselves (beyond the excluded volume interaction), a monolayer film is obtained \cite{Tadros1989, Evans1993, Elimelech1995, Rimai1996, Privman2000a, Privman2000b, Privman2000ed, Talbot2000, Schaaf2000, Cadilhe2007} and the model is characterized by an asymptotic jammed limit where no more adsorption events can possibly take place, and consequently a limiting value of the coverage is attained, i.e., the jamming coverage.

    \begin{figure}
      \includegraphics[width=6cm]{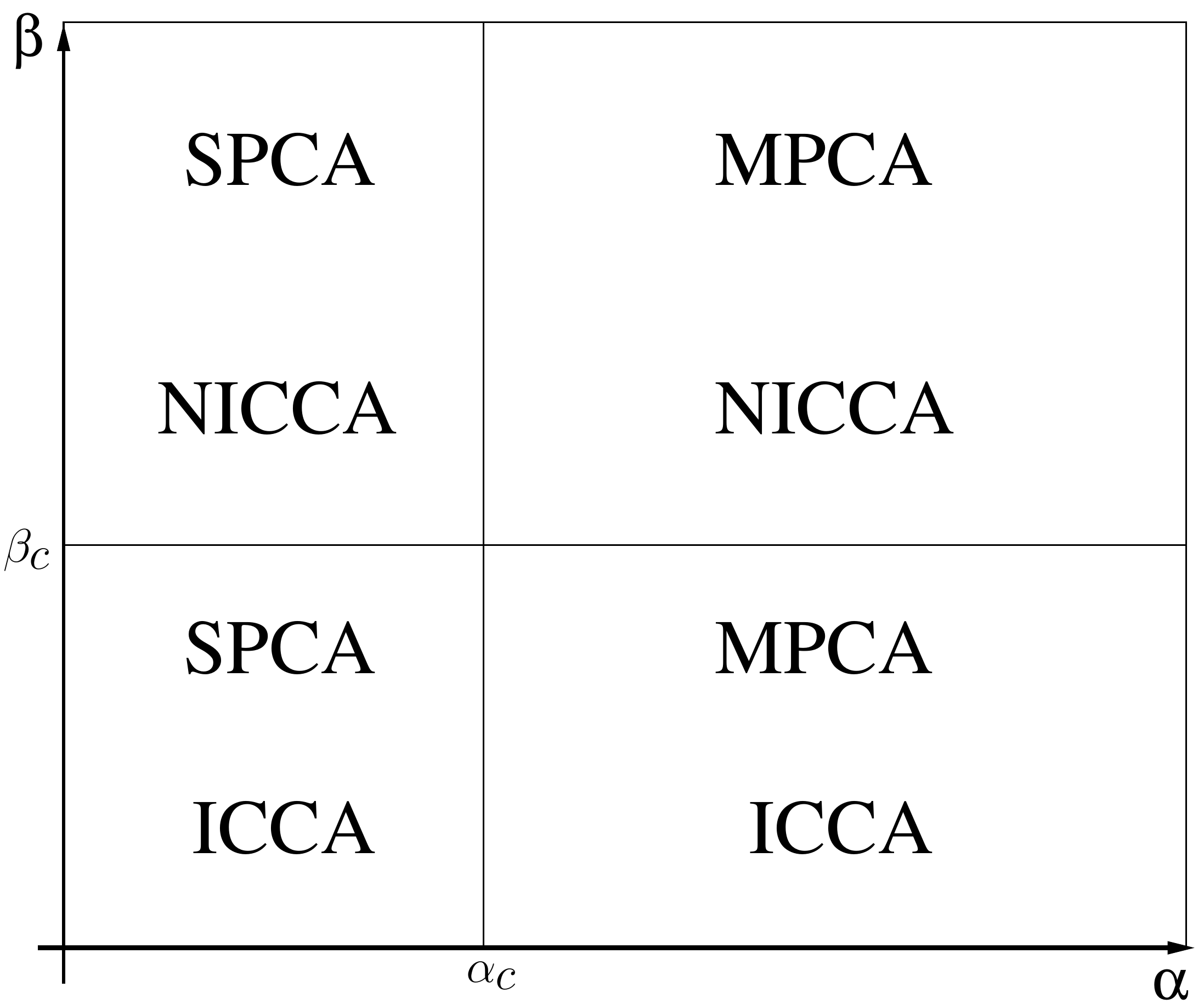}
      \caption{\label{fig:phase.diagram}
        Generalized version of the {\it phase diagram} for cell linear dimension $\alpha$ vs. inter-cell distance $\beta$.
        Cell-cell separation $\beta < \beta_c$ yields the \textit{interacting cell-cell adsorption\/} (ICCA) regime, while $\beta \ge \beta_c \ge 1$ gives the \textit{noninteracting cell-cell adsorption\/} (NICCA) regime.
        For cell sizes $\alpha < \alpha_c \le \sqrt {2}/2$ we find the \textit{single-particle-per-cell adsorption\/} (SPCA) regime, while for $\alpha \ge \alpha_c$ we find the \textit{multiparticle-per-cell adsorption} (MPCA) regime.
        }
    \end{figure}

    There is presently substantial literature on the adsorption of monodisperse colloids on regular substrates (i.e., without the presence of a pattern) \cite{Feder1980a, Evans1993, Privman2000a, Privman2000b, Talbot2000, Schaaf2000}.
    Here, we are primarily interested in addressing the effect of size dispersion on patterned substrates.
    Unfortunately, even such an extension poses several challenges that we partially address.
    As described above, the patterned substrate is flat with particles being allowed to adsorb inside the cell regions with a uniform random distribution.
    This is clearly an idealization as one expects different adsorption characteristics at the cell boundaries, but we do not address such effect in the present study.
    Furthermore, we take the colloidal particles to be disks.
    Strictly speaking, the particles cannot be modeled with full accuracy as disks since larger spherical particles can accommodate smaller spheres underneath them, and this effect was reported in the literature \cite{Talbot1989a}.
    Besides this effect, spheres of different radii have different overlapping rules than disks and even the meaning of what one considers as a cell, i.e., those regions of the substrate where adsorption can take place, becomes ambiguous.
    As an example of the latter, a pattern of cells in high relief would have different effective adsorption areas than one sunken relative to the interstitial cell space.
    In the case of sunken cells, particles of different radii attempting adsorption have different effective areas for adsorption.

    \begin{figure*}
      \begin{flushleft}
        {\bf\hspace*{0.75cm}(a)\hspace*{8.0cm}(b)}
      \end{flushleft}\vspace*{-0.1cm}
      \begin{tabular}{cc}
         \hspace*{0.0cm}\includegraphics[width= 8.0cm,height=8.0cm]{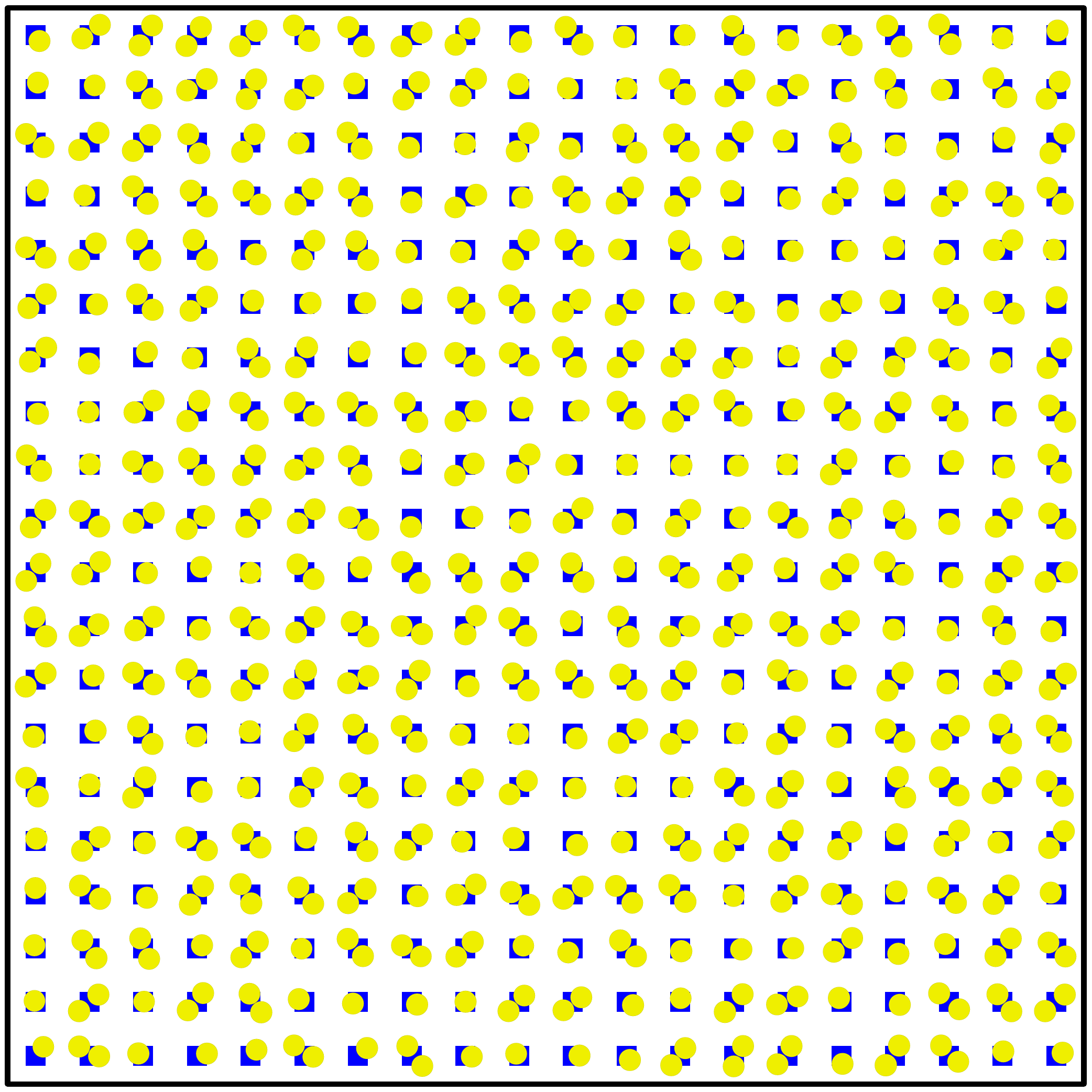}\hspace*{0.2cm}&\ \includegraphics[width=8.0cm, height= 8.0cm]{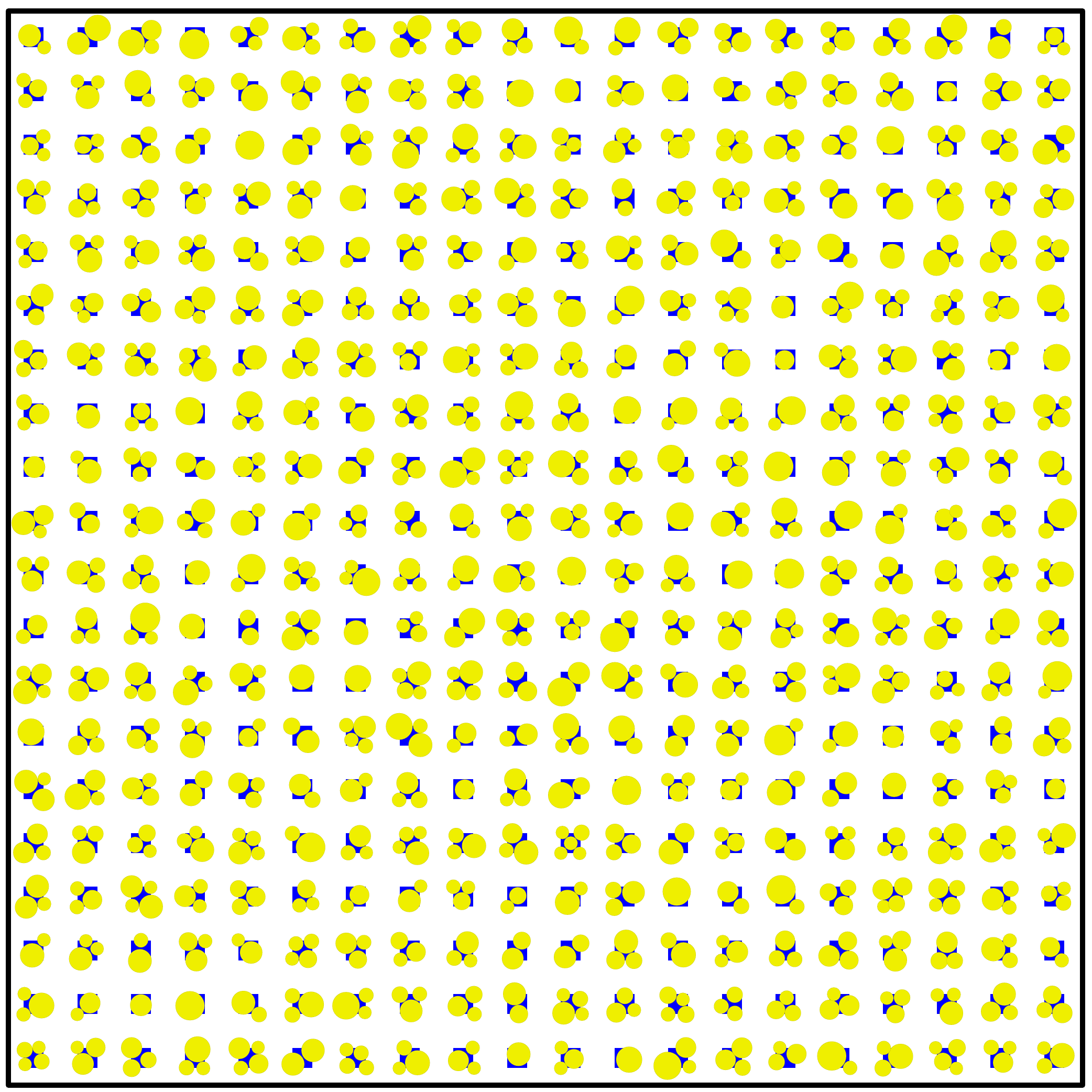}
      \end{tabular}
      \caption{\label{fig:snapshots}
      (Color online)
      Particles attempting adsorption have a truncated Gaussian-size distribution, as described in the text.
      The two snapshots represent $20 \times 20$ unit cells region from larger simulations showing a typical configuration at the jammed state with $\alpha= 0.96$.
      Cell-cell separation is large enough for the simulations to take place in the NICCA regime:
      At a size dispersion of $1\%$, part (a), $\beta= 1.5 > \beta_c= 1.02$, while for a size dispersion of $20\%$, part (b), $\beta= 1.5 > \beta_c= 1.4$.
      In (a), only monomers, dimers, and trimers within cells are possible (in fact, trimers are not observed in this snapshot), while in (b) tetramers and pentamers are also observed.
      }
    \end{figure*}

  \section{Some exact relations}\label{sec:results}
    We start by properly characterizing the various regions of a {\it phase diagram} as shown in Fig.~\ref{fig:phase.diagram}.
    The effect of a pattern has been previously studied in the context of the adsorption of monodispersed particles \cite{Cadilhe2007, Araujo2008}.
    As in the monodispersed particle case, we define two dimensionless parameters, namely, $\alpha$, as the linear dimension of the cell, and, $\beta$, as the distance between adjacent cells, both in units of the mean particle diameter, $2\langle r\rangle$.
    In terms of the inter-cell distance $\beta$, the diagram is divided in two distinct regions, namely, {\it interacting cell-cell adsorption} (ICCA) and {\it noninteracting cell-cell adsorption} (NICCA).
    In the former, adsorption of a particle at a given cell can be prevented due to overlap with a previously adsorbed one at a neighboring cell, while in the latter case, the distance between cells is too large for such an overlap condition to happen.
    Regarding the cell size, the value of $\alpha$ also splits the diagram in two regions, namely, one region where a single particle can be adsorbed, which we term {\it single-particle-per-cell adsorption} (SPCA), from another, where more than one particle can be adsorbed, which we term {\it multiparticle-per-cell adsorption} (MPCA).
    We are primarily interested in a range of cell sizes where the number of particles adsorbed per cell is one or, at most, a small number (less than six), since this provides the highest control on the actual number of particles per cell and on their sizes.

    A transition from ICCA to NICCA occurs when the distance between cells, $\beta_c$, is large enough to prevent the largest particles, i.e., those of radius $r_{\mbox{\tiny max}}$, from overlapping in the region between cells, i.e.,
    \begin{equation}\label{equ:beta}
      \beta_c= \frac{r_{\mbox{\tiny max}}}{\langle r\rangle}.
    \end{equation}
    Notice that $\beta_c \ge 1$ makes the transition to NICCA at a larger inter-cell distance than that of the monodisperse case.
    Now, the transition from SPCA to MPCA occurs when the linear dimension of the cell allows a second particle to adsorb in a close-packed configuration, which depends on the radius of the smallest particle, $r_{\mbox{\tiny min}}$.
    The critical value is
    \begin{equation}\label{equ:alpha}
      \alpha_c= \frac{\sqrt{2}}{2}\frac{r_{\mbox{\tiny min}}}{\langle r\rangle},
    \end{equation}
    with $\alpha_c \le \sqrt{2}/2$, i.e., at a cell size smaller than the monodisperse case.
    Only in the monodisperse limit one has $r_{\mbox{\tiny min}}= r_{\mbox{\tiny max}}= r \equiv \langle r\rangle$ and previously reported critical values are recovered~\cite{Araujo2008}.

    Notice that the values of $\beta_c$ and $\alpha_c$ separating the various regions of the {\it phase diagram} are independent of the actual functional dependence of the distribution function in terms of the particle radii.
    In fact, they only depend on the extreme values of the radii (minimum and maximum) as shown in Eqs.~(\ref{equ:beta}) and~(\ref{equ:alpha}) and the value of size dispersion.
    Now, if the limiting values of the radii are \footnote{The fact we take $\pm2\sigma$ represents just a choice, not a requirement.}
    \begin{equation}\label{equ:rmin}
      r_{\mbox{\tiny min}}= \langle r\rangle - 2\sigma
    \end{equation}
    and
    \begin{equation}\label{equ:rmax}
      r_{\mbox{\tiny max}}= \langle r\rangle + 2\sigma,
    \end{equation}
    then from Eqs.~(\ref{equ:beta})-(\ref{equ:rmax}), the critical values $\alpha_c$ and $\beta_c$ can be put in terms of the dimensionless size dispersion by defining the latter in terms of the mean radius as $\gamma= \sigma/\langle r\rangle$, so that
    \begin{equation}\label{equ:alpha2}
      \alpha_c= \frac{\sqrt{2}}{2}(1 - 2\gamma)
    \end{equation}
    and
    \begin{equation}\label{equ:beta2}
      \beta_c= 1 + 2\gamma,
    \end{equation}
    which allows a useful re-interpretation of Eqs.~(\ref{equ:beta}) and~(\ref{equ:alpha}), apart from also reducing the number of independent parameters to one.
    From Eqs. (\ref{equ:rmin}) and (\ref{equ:rmax}) one has $\gamma < 1/2$ \footnote{This is a direct consequence of considering symmetric values of $\sigma$ around the mean value.}, again due to the functional dependence adopted for $r_{\mbox{\tiny min}}$ and $r_{\mbox{\tiny max}}$ (Eqs.~(\ref{equ:rmin}) and~(\ref{equ:rmax}), respectively) in terms of $\sigma$.
    Now, taking two distinct distribution functions of the radii, but with the same size dispersion, Eqs. (\ref{equ:alpha2}) and (\ref{equ:beta2}) show that their {\it phase diagrams} are equivalent.

    \begin{figure}
      \includegraphics[width= 7.0cm]{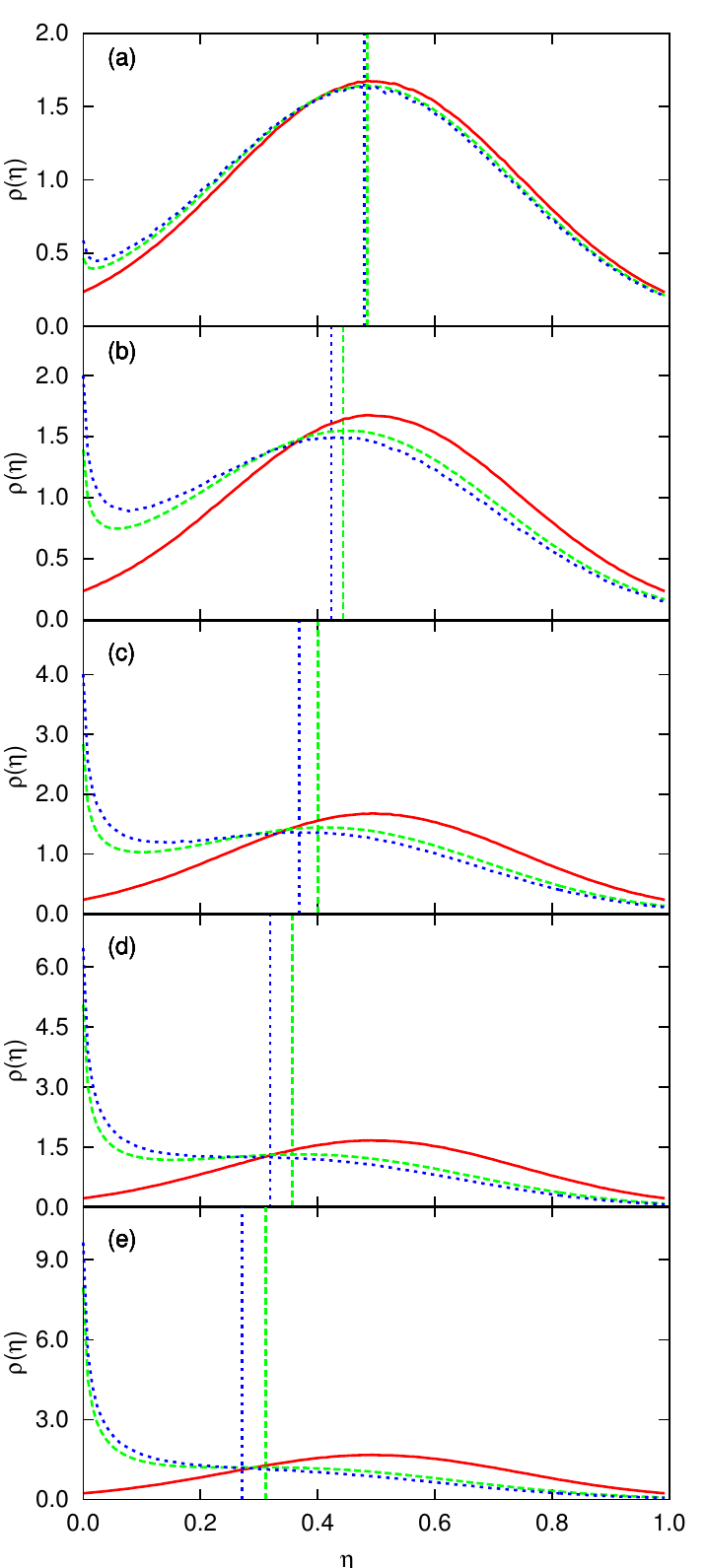}
      \caption{\label{fig:adsorbed.radius.alpha}
        (Color online) Adsorbed radii distribution for patterned surface with $\alpha$ equal to $0.4$ (solid, red line) and $1.7$ (dashed, green line), and the bulk case ($\alpha \to \infty$) (dotted, blue line).
        For size dispersion: a) $1\%$, b) $5\%$, c) $10\%$, d) $15\%$, and e) $20\%$.
        The vertical lines represent the mean normalized radius, $\langle\eta\rangle$, of adsorbed particles.
        Notice that for $\alpha= 0.4$ the corresponding value is $1/2$, regardless of the size dispersion.
      }
    \end{figure}

\begin{figure}
      \includegraphics[width= 7.5cm]{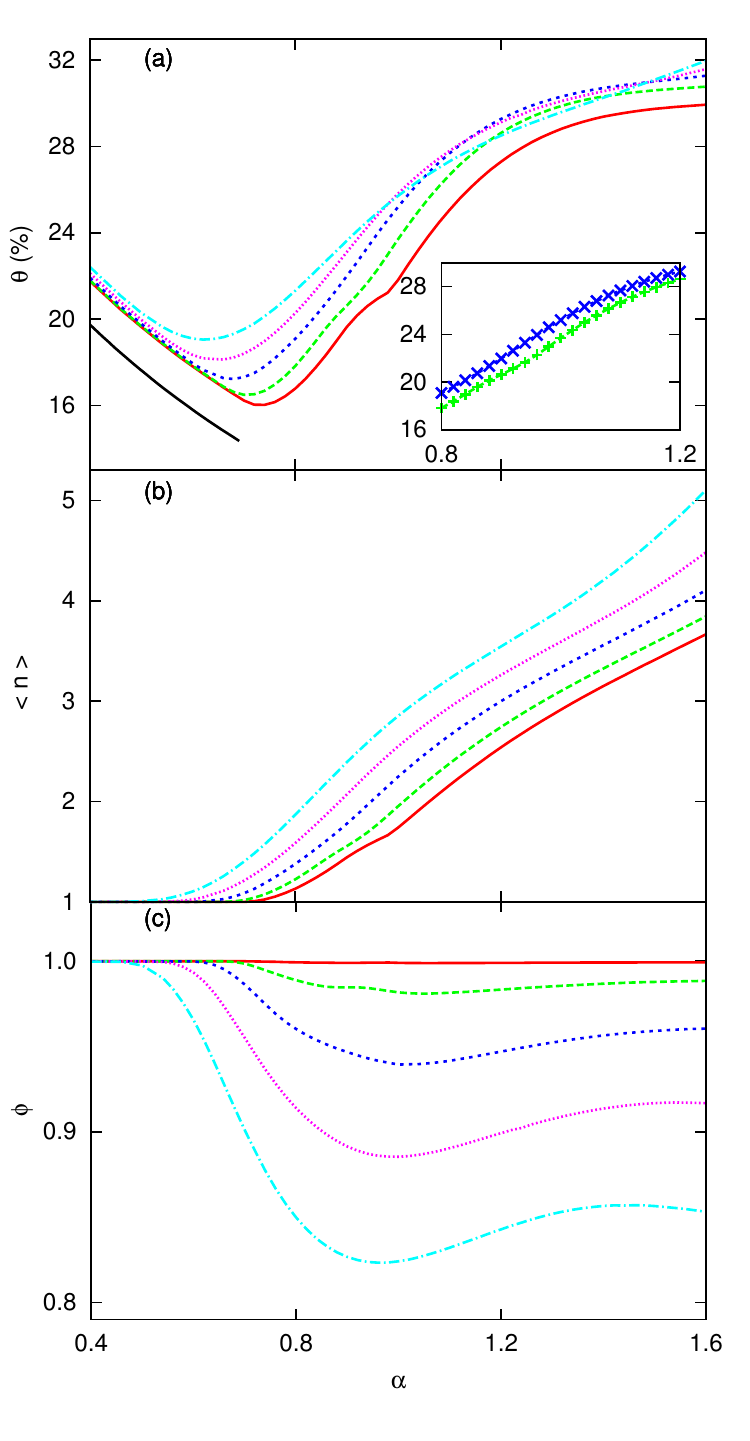}
      \caption{\label{fig:func.alpha}
      (Color online) For the {\it noninteracting cell-cell adsorption} regime, at the jammed state, as a function of the cell size: a) Coverage (for $\beta= 1.5$); b) Mean number of particles per cell; c) Mean adsorbed radius.
      Each line type corresponds to different values of the size dispersion, namely, solid (red) to $1\%$, dash (green) to $5\%$, dot (blue) to $10\%$, fine dot (purple) to $15\%$, and dash-dot (light blue) to $20\%$.
      In (a) the solid (black) line reproduces the monodisperse case in Ref.~\cite{Araujo2008} (vertically displaced for clarity).
      In the inset of (a) is magnified the region $0.8\leq\alpha\leq1.2$ for $5\%$ (lower line) and $10\%$ (upper line) of size dispersion (details in the text).
      The error bars are smaller than the line thickness.
      }
    \end{figure}

    In the present work, to systematize the effect of size dispersion on the adsorption process, attention is focused on the regime where cell-cell separation, $\beta$, is large enough so that adsorption on a given cell is not affected by particles previously adsorbed on neighboring cells since $\beta \ge \beta_c$ (NICCA regime).
    Depending on the value of $\alpha$ and the level of polydispersity of the particles, one can have one or more particles inside each cell.
    The effect of particle polydispersity is illustrated in Fig.~\ref{fig:snapshots} by snapshots of the jammed state for $\alpha= 0.96$ for two values of the size dispersion, namely, $\gamma= 1\%$ (Fig.~\ref{fig:snapshots}(a)) and $\gamma= 20\%$ (Fig.~\ref{fig:snapshots}(b))~\footnote{The actual definition of $\gamma$ is provided in the next section.}.
    For the remainder of the text we refer to values of the size dispersion in terms of their value relative to the mean radius.
    We denote as aggregates the set of adsorbed particles in a cell, and also name aggregates with a specific number of adsorbed particles, say, $1$, $2$, $3$, $\dots$ as monomers, dimers, trimers, $\dots$
    As in the monodisperse case, the inter-cell kinetics decouples so that one can follow the kinetics within each cell \cite{Araujo2008}.
    On the basis of this property we were able to extend exact relations obtained for the monodisperse case to account for size dispersion in both the single and multiparticle regimes.
    For example, in the SPCA regime, i.e., for $\alpha \le \alpha_c$,
the mean adsorbed radius becomes independent of both the cell size and
the actual distribution function of the radii, which equals that of the particles attempting adsorption.
    Consequently, the distribution of the radii of adsorbed particles follows a truncated Gaussian with the value of the dispersion also following that of the particles attempting adsorption.
    This can be better understood by considering the density distribution function of the normalized radii in the film, $\rho (\eta)d\eta$,
    where the normalized radius $\eta$ is defined as
    \begin{equation}\label{equ:eta}
      \eta= \frac{r - r_{\mbox{\tiny min}}}{r_{\mbox{\tiny max}} - r_{\mbox{\tiny min}}}.
    \end{equation}
    As defined, $\rho (\eta)$ represents the fraction of disks of normalized radius in the interval $]\eta, \eta + d\eta[$ such that
    \begin{equation}
      \int_0^1 \rho (\eta) d\eta= 1.
    \end{equation}
    In the SPCA regime, since the radii distribution of adsorbed particles follows that of the particles attempting adsorption, the mean normalized radius of the truncated Gaussian is $\langle\eta\rangle= 1/2$.
    Figure~\ref{fig:adsorbed.radius.alpha} illustrates these arguments in the particular case of $\alpha= 0.4$.
    In fact, it is possible to obtain an exact relation for the coverage in the SPCA regime, i.e., for $\alpha \le \alpha_c$, as
    \begin{equation}\label{equ:theta_J}
      \theta_J (\alpha, \beta, \gamma_{\mbox{\tiny T}})= \frac{\pi\left(1 + \gamma_{\mbox{\tiny T}}^2 \right)}{4(\alpha + \beta)^2} ,
    \end{equation}
    where $\gamma_{\mbox{\tiny T}}=\sigma_{\mbox{\tiny T}}/\langle r \rangle$, with $\sigma_{\mbox{\tiny T}}$ the effective size dispersion of particles attempting adsorption, i.e., $\sigma_{\mbox{\tiny T}}^2=\langle r^2\rangle-\langle r\rangle^2$.
    In deriving the above expression we considered the area occupied by a set of polydispersed particles, with only one particle per cell, relative to the area occupied by a monodisperse set of particles.
    Equation (\ref{equ:theta_J}) shows that in the polydisperse case the coverage is higher and that it increases by a factor of $1 + \gamma_{\mbox{\tiny T}}^2$ relative to the monodisperse case.
    Moreover, the coverage in the SPCA regime does not depend on the shape of the distribution function of the particle sizes.
    As an illustration, in Fig.~\ref{fig:func.alpha}(a) the simulation results, to be detailed in the next section, show the initial decay of the coverage proportional to $(\alpha + \beta)^{-2}$; deviations from this behavior coincide with the appearance of dimers (and larger aggregates), as shown in Fig.~\ref{fig:func.alpha}(b).
    The solid black line in Fig.~\ref{fig:func.alpha}(a) reproduces the monodisperse case found in Ref.~\cite{Araujo2008} (vertically displaced for clarity).

    Though it is not, in general, possible to calculate the jamming coverage for any value of the parameters $\alpha$, $\beta$, and $\gamma$ in the MPCA and NICCA regime, it is possible to relate the coverage to different values of $\beta \ge \beta_c$ by using
    \begin{equation}\label{equ:theta_J:relation}
      \frac{\theta_J (\alpha, \beta, \gamma)}{\theta_J (\alpha, \beta_o, \gamma)}= \left(\frac{\alpha + \beta_o}{\alpha + \beta} \right)^{\!\!2},
    \end{equation}
    where $\beta_o$ is the simulated value.
    The uncertainty for an ensemble of $N$ samples of the coverage is given by
    \begin{equation}
     \sigma_{\theta_J} (\alpha, \beta, \gamma)= \sqrt{\frac{\sum_{i= 1}^N \theta_{Ji}^2}{N} - \left(\frac{\sum_{i= 1}^N \theta_{Ji}}{N}\right)^{\!\!2}},
    \end{equation}
    and using Eq.~(\ref{equ:theta_J:relation}) one finds
    \begin{equation}\label{equ:sigma:theta_J}
      \frac{\sigma_{\theta_J} (\alpha, \beta, \gamma)}{\sigma_{\theta_J} (\alpha, \beta_o, \gamma)}= \left(\frac{\alpha + \beta_o}{\alpha + \beta} \right)^{\!\!2}.
    \end{equation}
    Equation~(\ref{equ:sigma:theta_J}) shows that the uncertainty of a particular simulation can be adjusted to the new $\beta$ value.
    Equations~(\ref{equ:theta_J})-(\ref{equ:sigma:theta_J}) extend similar equations for monodisperse particles \cite{Araujo2008} to the polydisperse particle case.
    In the following section, we analyze further results from extensive Monte Carlo simulations.

  \section{Monte Carlo simulations}\label{sec:mc.sim}
    We provide some motivation for a truncated exponential distribution of the size of particles attempting adsorption on a substrate, for which we foresee two possible scenarios.
    On one hand one can consider that the typical size dispersions observed in experiments are Gaussian distributed, and the fact that one does not observe all the possible sizes is mainly due to processing and the timescale taken for the observation.
    From a theoretical perspective, such approach poses difficulties on how to define the limiting coverage.
    Strictly speaking, for a Gaussian distribution one expects the coverage to approach unity, since particles of any size are possible, so the system does not jam, but this is not physically expected.
    On the other hand, taking a truncated exponential as the distribution function of the particle sizes has several benefits:
    it naturally accounts for the absence of the extreme values of the linear dimensions of the particles effectively present \cite{Meakin1992a, Adamczyk1997}, and removes the non-physical result of reaching a fully covered substrate.
    We define the truncated Gaussian-size distribution, $P(r)\mbox{d}r$, as
    \begin{equation}\label{equ:size.distrib}
      P(r)=
      \left\{
        \begin{array}{ll}
          A \exp\!{\left(-\frac{(r-\langle r\rangle)^2}{2\sigma^2}\right)}&\quad  r_{\mbox{\tiny min}} < r < r_{\mbox{\tiny max}}\\
          0&\quad\mbox{otherwise}\\
        \end{array}
      \right.
    \end{equation}
    where $\langle r\rangle$ is the mean radius, $\sigma$ is the dispersion as measured from a Gaussian distribution, and $A$ the normalization factor.
    All particles with radius below $r_{\mbox{\tiny min}}$ and above $r_{\mbox{\tiny max}}$ are discarded on the basis of being extreme values not actually observed within a typical experimental timescale \footnote{Further discussions can be found in \cite{Meakin1992a, Meakin1992b, Adamczyk1997}.}.
    The dispersions of the truncated Gaussian, $\sigma_{\mbox{\tiny TG}}$, and of the Gaussian, $\sigma$, are related by
    \begin{equation}
      f= \frac{\sigma_{\mbox{\tiny TG}}}{\sigma}=\left[1 - \frac{2^{3/2}}{\sqrt{\pi}e^2\mbox{erf} (\sqrt{2})}\right]^{1/2}.
    \end{equation}
    However, the actual cutoff values for $r_{\mbox{\tiny min}}$ and $r_{\mbox{\tiny max}}$ are given by Eqs.~(\ref{equ:rmin}) and (\ref{equ:rmax}), using the Gaussian distribution value, instead of the truncated Gaussian value, for a more direct interpretation.
    Consequently, the coverage in the SPCA regime is given by
    \begin{equation}\label{equ:theta_J:TG}
      \theta_J (\alpha, \beta, \gamma)= \frac{\pi\left(1 + f^2\gamma^2 \right)}{4(\alpha + \beta)^2}.
    \end{equation}

    To study the influence of cell size on various quantities such as the coverage, mean adsorbed radius, density of monomers, dimers, etc., and mean number of adsorbed particles per cell, we resort to extensive Monte Carlo simulations.
    From the experimental view, size dispersions above $20\%$ are rarely observed or of interest, so we consider size dispersions up to this value in the set of simulations \cite{Adamczyk1997}.
    As mentioned in Sec.~\ref{sec:model}, the transition from single to multiparticle adsorption occurs at $\alpha_c= (1-2\gamma)\sqrt{2}/2$ (Eq.~(\ref{equ:alpha2})).
    We also mention that $\beta= 1.5 > \beta_c= 1.4$ for the case of a size dispersion of $20\%$.
    Each simulation was carried out on a substrate of $500 \times 500$ unit cells and for an ensemble of $10^2$ samples.

    We used an efficient algorithm, to be described in greater detail in a separate publication~\cite{Araujo2010b}, while in the present work we present only an outline of its main aspects.
    We note that similar algorithms exist in the literature \cite{Meakin1992a, Privman1991a, Wang01_1994, Adamczyk1997, Cadilhe2007, Araujo2008, Araujo2010b}.
    The substrate is divided into a homogeneous mesh of squares and their size set to the minimum particle radius, $r_{\mbox{\tiny min}}$.
    We denote the squares as {\it mesh cells} to avoid confusion with the physically relevant cells of the pattern.
    Each mesh cell is classified as empty, occupied, or shadowed and only empty mesh cells are tested for adsorption.
    Occupied mesh cells contain the geometrical center of a particle and, therefore, they cannot be available for adsorption.
    Similarly, shadowed mesh-cells are those where adsorption cannot take place due to the excluded volume interaction of a previously adsorbed particle.
    Once the number of empty mesh-cells falls bellow a critical percentage, typically below $2\%$ of its initial value, the linear dimensions of the mesh cell are halved.
    Since mesh cells classified as {\it empty} can have part of their area shadowed, the fraction of shadowed area present in an empty mesh cell can be reduced by halving its linear dimensions and eliminating those offsprings that are now shadowed.
    On average, the procedure reduces the total shadowed area present in the empty mesh cells.
    Such a reduction of shadowed area leads to a higher probability of acceptance of a particle attempting adsorption, and, consequently, to improved algorithmic efficiency as compared to prior of the halving procedure.
    Further details of the algorithm are discussed in Ref.~\cite{Araujo2010b}.

    We start the analysis by considering the density distribution function of the normalized radii in the film, $\rho (\eta)d\eta$.
    At values of $\alpha \le \alpha_c$ the distribution of adsorbed radius follows that of the particles attempting adsorption.
    It must, thus, correspond to the truncated Gaussian distribution, and the observed distributions for $\alpha= 0.4$ in Fig.~\ref{fig:adsorbed.radius.alpha}(a)-(e) do corroborate the expectation.
    This is no longer the case for $\alpha= 1.7$ where several particles can fit inside a cell, but the smallest particle sizes, on average, have higher probability of adsorbing than larger ones.
    Though valid at all values of the size dispersion, this effect becomes more striking for the highest values of the size dispersion.
    As the values of the size dispersion increase $\rho (\eta)$ becomes more skewed towards smaller radii.
    The fact that particles of radii closer to $r_{\mbox{\tiny min}}$ are less probable affects the timescale for a successful adsorption to take place, since particles of larger radius cannot adsorb due to the overlap condition.
    The mean values of $\langle\eta\rangle$ are provided in Table~\ref{tab:eta}.
    Finally, the bulk case ($\alpha\to\infty$) has the highest skewness towards the smallest values of the radius; the mismatch between the distribution functions is larger (e.g., $\alpha= 1.7$) at small radii for the intermediate values of the size dispersion.
    Geometrical constraints imposed by small cell sizes effectively filters the smallest particles.
    The distribution function of the radii of adsorbed particles does show this trend with clearly asymmetric, non-Gaussian dependence on the normalized radius as show in Fig.~\ref{fig:adsorbed.radius.alpha}.
    Moreover, at a cell size of $\alpha= 1.7$, the distribution function of the adsorbed radii in the presence of a pattern closely follows that of regular (non-patterned) substrate particularly for the highly polydisperse case.

    \begin{table}[t]
      \caption{\label{tab:eta}Values of the mean normalized radius, $\langle\eta\rangle$ for the various values of $\sigma= 1\%$, $5\%$, $10\%$, $15\%$, and $20\%$ and the two values of $\alpha= 1.7$ and the bulk case ($\alpha \to \infty$) belonging to the MPCA regime as used in Fig.~\ref{fig:adsorbed.radius.alpha}.}
      \begin{tabular}{r@{\quad\qquad\qquad\qquad\qquad\qquad\qquad}l@{\qquad\qquad}l}
         \hline\hline\\
         $\sigma\ (\%)$&\multicolumn{2}{c}{\vspace*{1.0mm}$\langle\eta\rangle$}\\
         \cline{2-3}
         &\rule{0pt}{12pt}$\alpha= 1.7$&$\alpha \to \infty$\\
         \hline\hline\\
         1&0.4851&0.4792\\
         5&0.4437&0.4235\\
        10&0.4007&0.3686\\
        15&0.3569&0.3190\\
        20&0.3109&0.2712\\
        \hline\hline
      \end{tabular}
    \end{table}

    To further characterize the NICCA regime, we study the dependence of various quantities on the size of the cells, $\alpha$.
    We first consider the jamming coverage, $\theta_J$, as a function of the cell size, $\alpha$, for values of the size dispersion ranging from $1\%$ to $20\%$ as shown in Fig.~\ref{fig:func.alpha}(a).
    As observed in Sec.~\ref{sec:results}, values of the coverages, while within the SPCA regime, decay proportionally to $(\alpha + \beta)^{-2}$.
    Figure~\ref{fig:func.alpha}(a) shows the curve for the monodisperse case (offset down to $20\%$ instead of $21.76\%$ at $\alpha= 0.4$ and $\beta= 1.5$ for clarity) as a guide for similar behavior in the polydisperse cases, and, of course, one observes that coverage values increase as the mean number of particles per cell is no longer unit.
    In the MPCA regime, an increase of the jamming coverage with the cell size is observed, as shown in Fig.~\ref{fig:func.alpha}(a), regardless of the value of the size dispersion.
    For the highly monodisperse case of a size dispersion of $1\%$, the values of the jamming coverage with $\alpha$ closely follow those of the monodisperse case \cite{Araujo2008}, with features such as the cusp at the transition from up-to-two to up-to-three particles per cell present ($\alpha=(1 + \sqrt{3})/2^{3/2}\approx0.96$).
    More surprising is the presence of a faint signature of the cusp at a size dispersion of $5\%$ (see inset of Fig.~\ref{fig:func.alpha}(a)).
    Since the transition from SPCA to MPCA depends on the size of the smallest particle, an increase of the size dispersion leads to a lower transition threshold as shown in Fig.~\ref{fig:func.alpha}(a).
    This dependence can be observed in Fig.~\ref{fig:snapshots} for snapshots of the jammed state for size dispersions of $1\%$ (Fig.~\ref{fig:snapshots}(a)) and $20\%$ (Fig.~\ref{fig:snapshots}(b)) at a fixed $\alpha= (1 + \sqrt{3})/2^{3/2}$, which represents the transition from up-to-two to up-to-three particles per cell in a close packed arrangement in the monodisperse case.
    At this cell size and for a size dispersion of $1\%$, only monomers and dimers are observed (Fig.~\ref{fig:snapshots}(a)), though trimers could, in principle, be possible.
    At a size dispersion of $20\%$, larger aggregates become possible, like pentamers (Fig.~\ref{fig:snapshots}(b)).

\begin{figure}
      \includegraphics[width= 7.0cm]{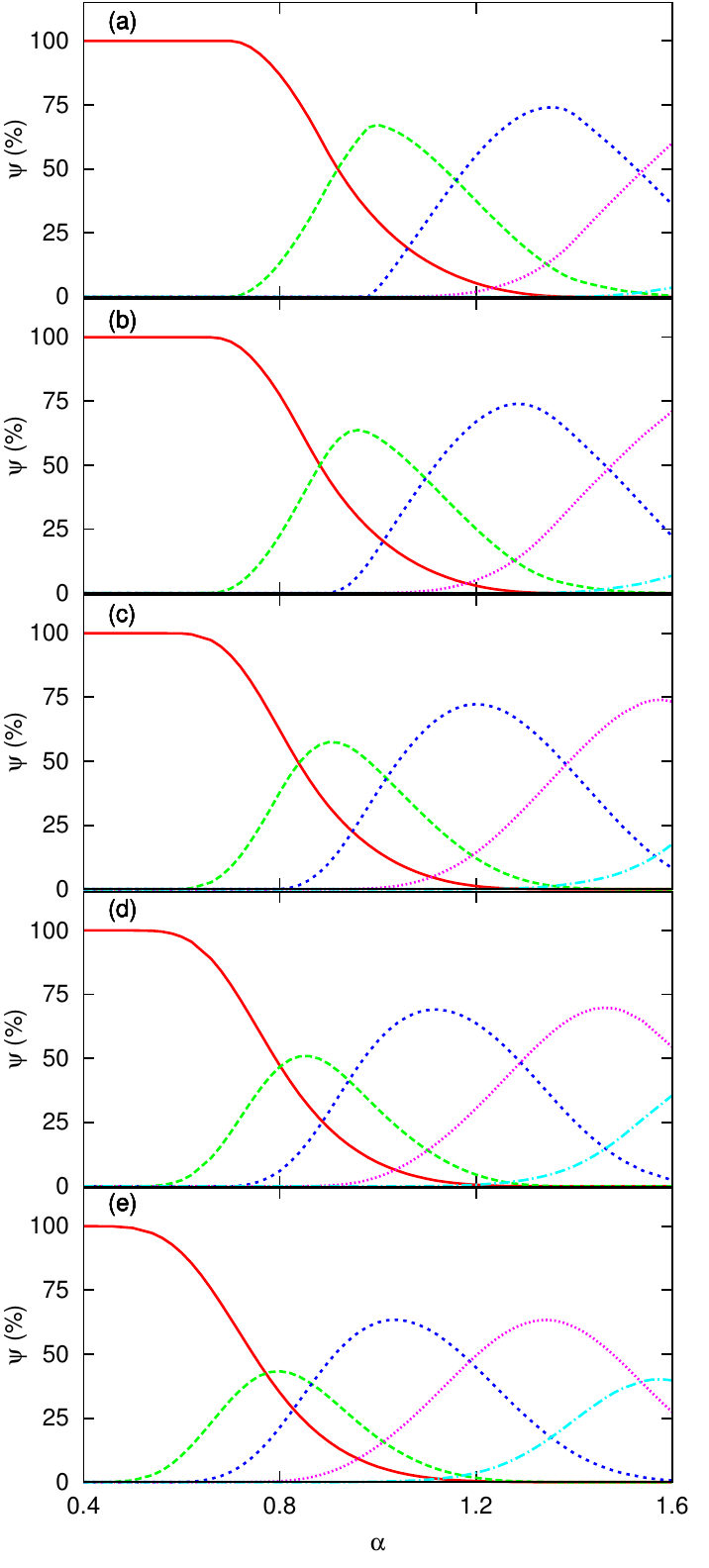}
      \caption{\label{fig:npercell.alpha}
      (Color online) Density of monomers (red, solid line), dimers (green, dashed line), trimers (blue, dotted line), tetramers (magenta, fine dot line), and pentamers (light blue, dash-dot line) for different values of the size dispersion: a) $1\%$, b) $5\%$, c) $10\%$, d) $15\%$, and e) $20\%$.
      }
    \end{figure}

    In general, the higher the values of the size dispersion, the higher the coverage, at the expense of a lower control over both the number of particles adsorbed on each cell and their sizes.
    The mean particle number per cell is shown in Fig.~\ref{fig:func.alpha}(b).
    The mean adsorbed radius, $\phi$, of the resulting film as a function of the cell size is shown in Fig.~\ref{fig:func.alpha}(c).
    The quantity reflects the balance between the population of {\it small} and {\it large} adsorbed particles as a function of the cell size and the degree of polydispersity of the particles.
    At small size dispersions, the values of the radii remain relatively undifferentiated with nearly symmetric distributions as shown in Fig.~\ref{fig:adsorbed.radius.alpha}(a) for the case of $1\%$.
    As the size dispersion increases, the distribution of the adsorbed radii becomes more skewed towards smaller values.
    The overall result is reflected by a substantial decrease of the average value of the mean adsorbed radius (at fixed $\alpha$) as a function of the size dispersion as shown in Fig.~\ref{fig:func.alpha}(c).
    This can be illustrated by considering the case of dimers for values of $\alpha$ above, though close, to $\alpha_c$.
    In this limit only small particles are allowed to adsorb in the formation of dimers, so the mean adsorbed radius must decrease.
    However, at a fixed value of $\alpha$, say $0.8$, and for increasing values of the size dispersion, the relative population of particles changes significantly as shown in Fig.~\ref{fig:npercell.alpha}.
    As the value of $\alpha_c$ shifts to lower values with increasing values of the size dispersion, dimers and trimers become possible.
    At the transitioning points, from monomer to dimer and from dimer to trimer, the cell acts as a filter to allow adsorption of the smallest particles, i.e., particles with radii at or close to $r_{\mbox{\tiny min}}$.
    As the values of $\alpha > \alpha_c$ increase the fraction of dimers grows, peaks, and decays; this feature reproduces itself on larger particle aggregates like trimers, tetramers, etc.
    The distribution functions are skewed towards the large cell sizes.
    At large values of cell size ($\alpha>1$), although the mean number of particles per cell increases with the size dispersion (Fig.~\ref{fig:func.alpha}(b)), yet the decrease in the mean adsorbed radius is such (Fig.~\ref{fig:func.alpha}(c)) that the coverage is no longer a monotonic increasing function of the size dispersion. 

    Finally, we briefly address possible particle arrangements inside a cell.
    For example, dimers tend, on average, to orient themselves along the diagonals of the cells.
    Considering cells sizes that accommodate up to three particles, one can understand the underlying reasons.
    Particles adsorbing at the center may block further particles to adsorb and become monomers.
    Now, if two particles adsorb along the diagonal, this blocks the possibility of a third particle to adsorb, thus becoming a dimer.
    Finally, two particles adsorbed along an edge permit a third particle to fit in and form a trimer.

  \section{Concluding remarks}\label{sec:conclusion}
    We performed extensive Monte Carlo simulations of the irreversible adsorption of polydisperse disks on a patterned substrate.
    A pattern consisting of square cells positioned at the vertices of a square lattice was considered, but extensions and generalizations to other lattice arrangements and cell shapes are straightforward.
    We used a physically motivated truncated Gaussian-size distribution to model the polydispersity of the particles attempting adsorption and size dispersions up-to $20\%$ of the average radius.
    The model extends a previous study of the irreversible adsorption of monodispersed particles on patterned substrates~\cite{Araujo2008}.
    In the present work, we focused on the noninteracting cell-cell adsorption regime (NICCA) and on the jammed state properties.

    The model is suitable to describe relevant features of colloidal particle adsorption under the assumption that no particle can fit under another particle of larger radius as discussed in Sec.~\ref{sec:model}.
    Even though the modeling relies on the excluded volume interaction, several quantities are generally valid, even if more realistic interactions are taken into account, like Coulomb and van der Waals.
    For example, the various transition points of the {\it phase diagram}, as, e.g., the transition from monomer to dimer solely depends on geometrical parameters such as the minimum and maximum values of the radii of particles attempting adsorption.
    Hence, these transition values are not affected by the inclusion of more complex interactions especially for the more interesting case of small cell sizes, where geometrical constraints are more significant and one observes substantial departure from the bulk (regular) substrates.
    In contrast, the interparticle distance of the adsorbed particles will depend on the choice of interaction potential, or the presence of hydrodynamic effects, since the quantity depends on the way adsorbed particles interact with a landing particle.
    Coverage represents another quantity that depends on the interactions taken into account.

    Coverage efficiency, for the excluded volume interaction case, improves with size dispersion even accounting for the fact that the adsorption of small particles is favored.
    The presence of a pattern favors adsorption of small particles for cell sizes near, but above, the transition points, e.g., from monomer to dimer and dimer to trimer.

    Finally, the present study provides insight on the values of the parameters required to tune the average number of particles adsorbed per cell.

  \begin{acknowledgments}
    A.C. thanks both Funda{\c c}{\~a}o para a Ci{\^e}ncia e a Tecnologia (SFRH/BPD/34375/2007) and Funda{\c c}{\~a}o Calouste Gulbenkian for fellowships to visit Los Alamos National Laboratory and acknowledges the warm hospitality of the T-1 Group at Los Alamos National Laboratory.
    We thank Vladimir Privman and Cynthia Reichhardt for useful comments on the manuscript.
  \end{acknowledgments}
  \bibliography{rsa,book}
\end{document}